# The expression of ensemble average internal energy in long-range interaction complex system and its statistical physical properties

Yanxiu Liu[1], Shenglei Zhang[1], Liu He[1], Cheng Xu[1], Zhifu Huang[1,*]

1.College of Information Science and Engineering, Huaqiao University, Xiamen 361021, People's Republic of China

In this paper, we attempt to derive the expression of ensemble average internal energy in long-range interaction complex system. Further, the Shannon entropy hypothesis is used to derive the probability distribution function of energy. It is worth mentioning that the probability distribution function of energy can be equivalent to the q-Gaussian distribution given by Tsallis based on nonextensive entropy. In order to verify the practical significance of this model, it is applied to the older subject of income system. The classic income distribution is two-stage, the most recognized low-income distribution is the exponential form, and the high-income distribution is the recognized Pareto power law distribution. The probability distribution can explain the entire distribution of United States income data. In addition, the internal energy, entropy and temperature of the United States income system can be calculated, and the economic crisis in the United States in recent years can be presented. It is believed that the model will be further improved and extended to other areas.

*Email: zfhuang@hqu.edu.cn

# 1 Introduction

In long-range interaction complex systems, the interaction energy of the system can not be ignored. Thus, the way to calculate the interaction energy of the system is really important. In order to consider this problem, we borrow the concept of the ISING model [1]. The ISING model assumes that the interaction between two first neighbor particles is proportional to the product of their respective rotations, which gives us a great deal of inspiration. Unlike the ISING model, we assume that the interaction of two arbitrary systems is proportional to the product of their respective energies, starting from the interaction of the two systems, the ensemble average internal energy of N systems and the probability distribution function are derived. After that, the established model is applied to the income system. Using typical United States income data, the cumulative income probability distribution of the United States can be fitted with two parameters, and the corresponding statistical physical properties of the income system are associated with actual economic and social development. Therefore, this paper has two main starting points: First, a simple model of long-range interaction complex system is established. As long as the conditions are established, it can be applied to other disciplines. Second, the statistical physical properties of the income system are analyzed. The related statistical physical quantity can better describe economic fluctuations, which may allow scholars to look at the income system from another angle.

In addition, Pressé et al. [2] summarized the modeling method of non-exponential distribution such as power law, and considered that non-expansibility should be represented by constraints. Entropy may not be a usable constraint because it has a wide range of effects and may not be related to the energy allocation between systems. Therefore, our work can enable researchers to examine the issue of a long-range interactionl system from another new perspective and lay the groundwork for future work.

**2 Model**

Suppose the internal energy of an isolated system is E, which is composed of two subsystems. The internal energy of the two subsystems can be expressed as $u_1, u_2$, which will change over time due to energy exchange. If the interaction between them can be ignored, then E is a simple addition of the internal energy of the two single subsystems. However, in the real world, the interactions between subsystems should not be ignored. Although interaction energy usually, in theory, is negligible due to the complexity of interactions. Inspired by the ISING model, we assume that the interaction between subsystems is as follows $\lambda u_1 u_2$. It is worth noting that the assumptions we take are considered to be the simplest and do not include many factors, such as distance. Also, we introduced a quantity $\tilde{u}$, which is defined $u_1$ equals to $u_2$, and we name $\tilde{u}$ as ensemble average internal energy. Consequently, the system energy E is written as:

$$E = u_1 + u_2 + \lambda u_1 u_2 = \tilde{u} + \tilde{u} + \lambda \tilde{u} \tilde{u} \tag{1}$$

Multiply the two sides of the equation by $\lambda$ and plus 1, we can get

$$1 + \lambda E = (1 + \lambda u_1)(1 + \lambda u_2) = (1 + \lambda \tilde{u})^2 \tag{2}$$

By definition, the ensemble average internal energy is equal to

$$\tilde{u} = (\sqrt{1 + \lambda E} - 1) / \lambda \tag{3}$$

The internal energy $u_1(t)$, $u_2(t)$ of each system evolve over time (for example, the exchange of internal energy), and the arithmetic average $\overline{u(t)} = [u_1(t) + u_2(t)]/2$ also evolves over time, yet the ensemble average internal energy $\tilde{u}$ is a conserved quantity related to the total energy E and interacting coefficient $\lambda$.

If there are three systems in this system, there are two ways to write the total. In case 1, a pairwise interaction i.e.

$$E = u_1 + u_2 + u_3 + \lambda u_1 u_2 + \lambda u_2 u_3 + \lambda u_1 u_3 \tag{4}$$

In case 2, regarding the subsystems $u_1$ and subsystems $u_2$ as a whole, their overall energy is $u_{1+2} = u_1 + u_2 + \lambda u_1 u_2$. Consequently, in this case, the internal energy of the system can be written into the following formula.

$$E = u_{1+2} + u_3 + \lambda u_{1+2} u_3 = u_1 + u_2 + u_3 + \lambda(u_1 u_2 + u_2 u_3 + u_1 u_3) + \lambda^2 u_1 u_2 u_3 \tag{5}$$

Comparing the above two cases, one can find that the three subsystems' interaction energy $\lambda^2 u_1 u_2 u_3$ has been considered in case 2.

When calculate the internal energy with Eq.(5), the three subsystems' interaction energy $\lambda^2 u_1 u_2 u_3$ should be joined in case 1. As a result, we are obliged to introduce an interaction coefficient of three subsystems and its magnitude is $\lambda^2$. In general, all possible interactions among multi-system must be taken into account. Therefore, the total internal energy of the three-system system should be the Eq.(5), Multiply the two sides of the equation by $\lambda$ and plus 1, and use the three-system ensemble average internal energy $\tilde{u}$ to represent:

$$1 + \lambda E = (1 + \lambda u_1)(1 + \lambda u_2)(1 + \lambda u_3) = (1 + \lambda \tilde{u})^3 \tag{6}$$

It can be seen that the ensemble average of internal energy of the three systems is only related to the total internal energy of the system. By using the same method, $\lambda_4$, $\lambda_5, \ldots, \lambda_N$ can be derived and $\lambda_m = \lambda^{m-1}$ can be obtained. Based on the discussion all above, the internal energy of system can be written as E= single subsystem's energy + two subsystems' interaction energy + three subsystems' interaction energy +…+ N subsystems' interaction energy. Thus, the internal energy of the system with N interaction subsystems can be derived as

$$E = \sum_i u_i + \lambda_2 \sum_{i,j(i \neq j)} u_i u_j + \lambda_3 \sum_{i,j,k(i \neq j \neq k)} u_i u_j u_k + ,\ldots, + \lambda_N \prod_i u_i \tag{7}$$

where $\lambda_N = \lambda^{N-1}$. Note that Eq. (7) can be simplified as

$$1 + \lambda E = \prod_i (1 + \lambda u_i) = (\lambda \tilde{u} + 1)^N \tag{8}$$

Where $\tilde{u}$ is ensemble average internal energy. Taking logarithms and $\tilde{u}$ can be given

$$\tilde{u} = \{\exp[\sum_i P_l \ln(1 + \lambda u_l)] - 1\} / \lambda \tag{9}$$

where $P_l$ is the probability distribution function of a subsystem with energy $u_l$.

In addition, we take the form of Shannon entropy [3] that is universally applicable in various scientific fields.

$$S = -k \sum_l p_l \ln p_l \tag{10}$$

By using the method of Lagrange's undetermined multipliers

$$\delta(\frac{S}{k} - \alpha \sum_l p_l - \beta \tilde{u}) = 0 \tag{11}$$

and substituting the Eq. (9) and Eq. (10) into Eq. (11),

$$P_l = e^{-1-\alpha}(1 + \lambda u_l)^{-\gamma} \tag{12}$$

Where $\gamma = \beta \exp[\sum_i P_i \ln(1+\lambda u_i)]/\lambda$. Eq.(10) is actually equivalent to the q-Gaussian distribution: $p(u) = [1-(1-q)\beta_T u]^{1/(1-q)}/Z_T$, where $q = 1/\gamma + 1$. In addition, the cumulative probability distribution function can be derived by Eq. (12):

$$C(u) = (1+\lambda u)^{1-\gamma} \qquad (13)$$

From the above theoretical model, it can be seen that the ensemble average internal energy (hereinafter referred to as internal energy), entropy, is only related to two parameters $\lambda$ and $\beta$. Therefore, we use the numerical processing method to accurately draw out the relationship between $\tilde{u}$, $S$ and $\lambda$, $\beta$ respectively, as shown in Figure 1.

From Figure 1, it shows that the internal energy and entropy increase with increasing temperature, which is in accordance with the second law of thermodynamics. Entropy increases spontaneously as the system moves from imbalance to equilibrium. it also shows that when the temperature tends to 0, the graph appears dashed, and each parameter will have an imaginary number at this time. From another perspective, in the real world, temperature cannot reach absolute zero. In addition, in practical applications we get in general, $\beta/\lambda$ is more than 2. It can be seen from the $\tilde{U}$ - $\lambda$ relationship diagram that when $\lambda$ becomes larger, $\tilde{U}$ will obviously increase, which is in accordance with the law of this model. And as $\lambda$ becomes larger, entropy also increases.

## 3. Data analysis

Statistical Physical Model is universally applied by other scientific field. After establishing the basic model of long-range interaction complex system, other scientific fields could be applied. So, this paper applied this model to the income system. It is necessarily to research on the probability distribution function of money, wealth and income in the society, which overlaps with long-term economic studies of social inequality. An economist called Pareto points out that the probability distribution of cumulative income in society is a power law function [4]. However, the actual data proves that the power law only applies to the upper end of income distribution. Fitting United States Revenue Data by Yakovenko and Rosser, it confirms that American society has a clearly defined two-level structure [5]: the overwhelming majority of the population (middle and low income population) subjects to exponential distribution and the other (high-income population) subjects to the power law distribution. Besides, Yakovenko and Rosser advance a theory based on thermal equilibrium of statistical mechanics to solve the exponential income distribution which received more and more empirical research support [6-10]. Also worth noting is Banerjee and Yakovenko believe that addition and multiplication can coexist, and an income distribution formula with three parameters is deducted by using Foker-Plank Equation [11]. Then a group of followers used the same or similar method to fit the whole income distribution. They have not established a statistical physical model to analyze the entire income distribution. The physical model we offer is not only applicable to the whole social stratum, but also fitting the whole income with two parameters, which is never been realized by other researchers.

The latest data of the distribution of personal income in the United States tax returns for the period 1995-2011 prepared by the Internal Revenue Service (IRS) [12]. The data is fitted with the derived cumulative probability distribution Eq. (13) showing in Figure 2. It can be seen that the formula derived can basically fit the income of the United States. When fitting the data of each year, the value of $\lambda$ each year can be obtained, and the value of $\beta$ each year can be calculated (corresponding to temperature, $\beta = \frac{1}{kt}$), together with the internal energy and entropy. Then the evolution of the above parameters in the United States income system over time can be drawn in Figure 3.

As we all know, the economy of the United States was very unstable around 2000. Since April 2000, the Internet economic bubble has begun to burst, and a large number of United States Internet companies have gone bankrupt. In 2001, the '9.11' incident led to a large amount of international funds being invested in the euro, and the United States faced large-scale capital flight. In August 2008, United States housing stock prices plummeted, the financial crisis swept the world, and the financial turmoil had early signals in April 2007. Applying this model to the income system, we can see that the occurrence of the economic crisis has indeed led to the corresponding changes in the relevant physical quantities.

From 1997 to 2000, the stock price of the United States dollar continued to rise, especially the Asian financial turmoil from 1997 to 1998 caused a large amount of funds to flow into the United States, the $\lambda$ on the absolute value of slope of time increase significantly, but Internal energy, entropy, temperature rise evenly, it means the United States economy is at a relatively rapid growth stage. After 2000, due to the

Internet economic crisis and the '9.11' incident, the economy was hit, Growth can no longer be as fast as it used to be, at this time, the value of $\lambda$ drops rapidly and forms a depression before 2006. The internal energy and temperature only increased slightly before 2002, almost unchanged, and the entropy decreased slightly. Before the economic bubble in 2007, the United States government recovered after a series of policies, but the financial crisis erupted in August 2008. As can be seen from figure, the absolute value of the $\lambda$ slope in 2007 suddenly increased, and its height is even higher than the Internet economic crisis. And $\lambda$ once again formed a depression before 2010, internal energy, entropy and temperature have a minimum in 2009. After that, the growth rate slowed significantly. But overall, $\lambda$ shows a downward trend, and internal energy, temperature, and entropy show an upward trend. This may indicate that, despite the economic crisis, the average level of people's lives is indeed better than before. In short, the change of the $\lambda$ may represent the development of the economy. Its sunken or bulging processes represent the whole process of economic crisis, from omens to outbreaks to recovery; while the internal energy, entropy and temperature represent the result of economic crisis.

What can be seen from the above analysis is: Entropy has been increasing almost all the time, and it has remained almost constant or declined slightly during the economic crisis, which is in line with the second law of thermodynamics: In a closed system, the entropy is always increasing from the unbalanced to the equilibrium spontaneous process, and remains unchanged in extreme cases. The overall trend of $\lambda$ is also decreasing year by year in addition to the abnormal reduction in the economic crisis, which shows that this model can describe the statistical physical properties of complex systems with long-range interactions without violating the axioms.

## 4. Summary

In conclusion, First, in the process of theoretical derivation, we did not ignore the interaction between subsystems, and the theoretical results are more consistent with the actual situation. Second, Theoretical models can be described simply, and only two parameters are needed to fit the cumulative income probability distribution, which has not been achieved in the previous work of researchers. Third, it can fit the income data of the whole period well, no matter it is ultra-low income, middle and low income or high income class. Fourth, the model is rich in content, and the relevant statistical physical quantities of the income system can be obtained every year, and the changes of these quantities can be related to the actual economic fluctuations. It is believed that the model is worthy of further study and extension to other fields.

## Acknowledgments

Project supported by the National Natural Science Foundation (No. 11775084,11305064).

**Figure captions:**

FIG．1． Statistical physical quantity changes with parameter temperature and interaction coefficient

FIG．2． Cumulative probability income fit map of the United States from 1995 to 2011

FIG．3． The statistical physical quantity changes with the year.

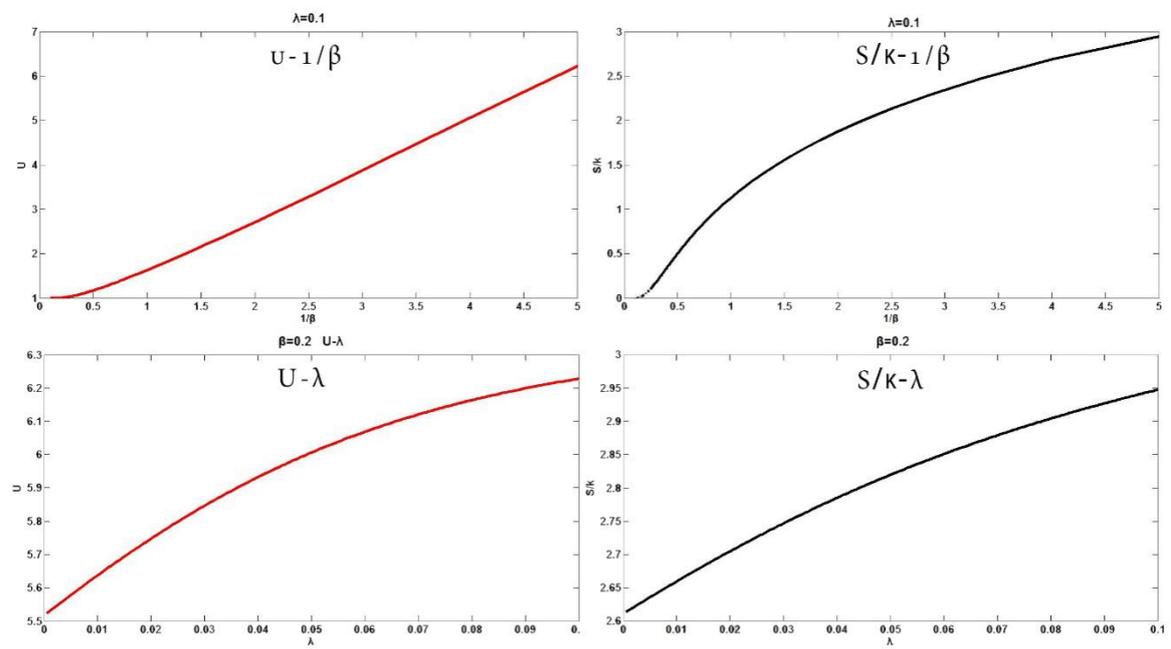

Fig. 1

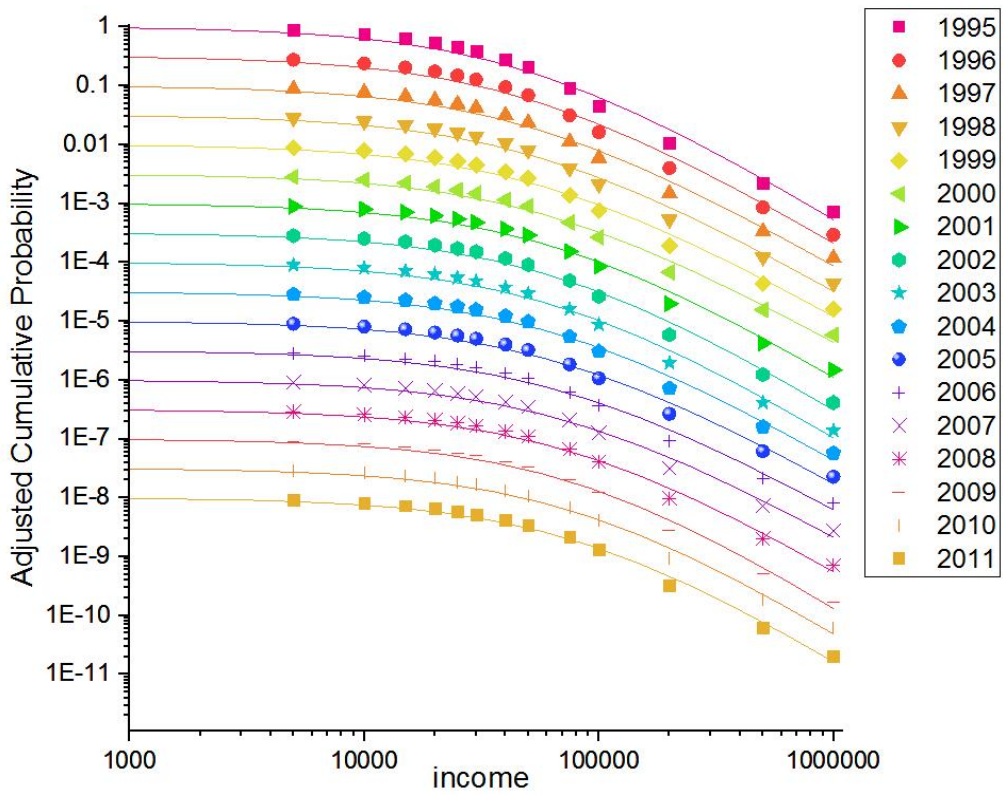

Fig. 2

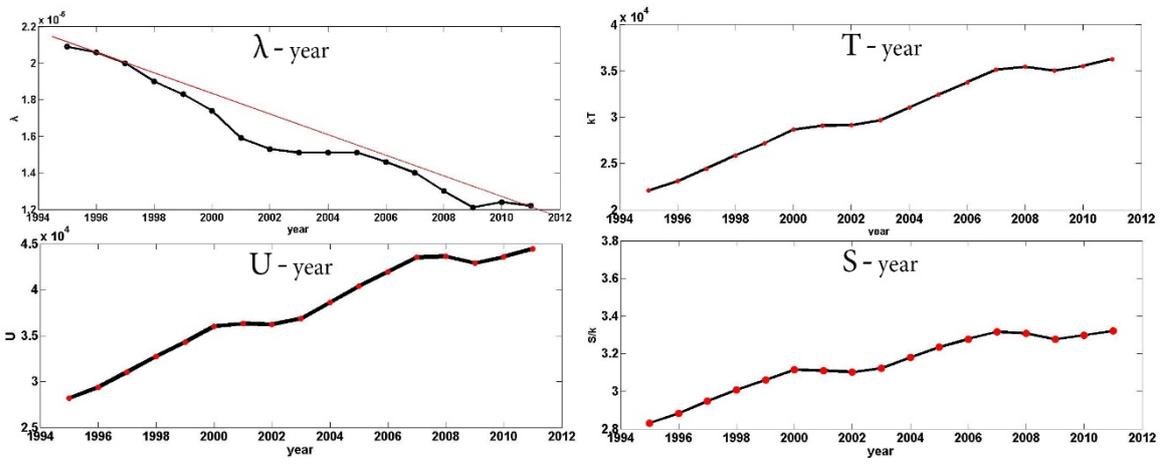

Fig. 3